\documentclass[preprint2]{aastex}
\usepackage{amsmath}
\usepackage{url}
\usepackage[T1]{fontenc}

\shorttitle{A $BVR_CI_C$ Survey of W Ursae Majoris Binaries}
\shortauthors{Terrell et al.}

\begin{document}

\title{A $BVR_CI_C$ Survey of W Ursae Majoris Binaries}

\author{Dirk Terrell}
\affil{Department of Space Studies, Southwest Research Institute, 1050 Walnut St., Suite 300,
    Boulder, CO 80302}
\email{terrell@boulder.swri.edu}

\author{John Gross}
\affil{Sonoita Research Observatory, 1442 E Roger Rd, Tucson, 85719}
\email{johngross3@msn.com}

\and

\author{Walter R. Cooney, Jr.}
\affil{Lot No. 23, Palanamai Village, 281 Burapha Golf Club, Moo 4, Bung, Sriracha, Chonburi 20110 Thailand }
\email{waltc@cox.net}

\begin{abstract}
We report on a $BVR_CI_C$ survey of field W Ursae Majoris binary stars and present 
accurate colors for 606 systems that have been observed on at least
three photometric nights from a robotic observatory in southern Arizona. Comparison with earlier 
photometry for a subset of the systems shows good agreement. We investigate two independent 
methods of determining the interstellar reddening, although both have limitations that can render
them less effective than desired. A subset of 101 systems shows good agreement between the two reddening 
methods.
\end{abstract}


\section{Introduction}

Overcontact binary stars\footnote{We note that there is a preference among some in the community for the term "contact" rather than "overcontact." \cite{rucinski97a} and \cite{wilson01} discuss both sides of the issue.} are systems in which both components share a common envelope between the inner and outer contact surfaces, and are therefore in physical contact with one another, leading to luminosity and mass exchange. Such systems show light curves with very similar eclipse depths and little color variation over the orbital cycle. The W Ursae Majoris systems are the most well-known overcontact systems, consisting of F, G and K-type stars with convective common envelopes. But early-type systems with radiative common envelopes do exist \citep{terrell03} and are even quite common among the O-type stars \citep{hilditchbell87}.

Despite over four decades of efforts to understand their internal structure and evolution, no model has emerged as clearly superior in explaining their observed properties, although active work (e.g., \cite{stepien09} and \cite{stepien11}) does show promise. Almost all of the theoretical work on overcontact systems has involved the W UMa systems, although the evolutionary models of \cite{nelsoneggleton01} show that the overcontact state is achieved for a wide variety of binary initial conditions (\it viz. \rm their Figure 9). The seminal work by \cite{lucy68} showed that overcontact systems with common convective envelopes could exist at zero-age but the agreement of his models with the observed period-color relation \citep{eggen67} was not very good and Eggen's period-color relation has played a critical role in the testing of models of W UMa systems ever since. The primary goal of this survey is to provide a significantly larger sample of overcontact systems with accurate photometry and to extend the photometry to the $R_CI_C$ passbands. 

Although the proliferation of relatively inexpensive CCD cameras and telescopes has resulted in a large body of light curve data for W UMa systems, most of these data are instrumental differential magnitudes rather than transformed standard magnitudes, a circumstance dictated in most cases by the predominantly non-photometric skies available to the observers. \cite{rucinski06} discusses the period-color-luminosity relationship for W UMa systems and its use as a distance indicator for these systems. One goal of our program is to provide more accurate color data than is currently available for this calibration. Accurate colors are also necessary for light curve solutions where the spectral type of the system is not available. 

\section{Sonoita Research Observatory}
\subsection{Hardware}

The goal of our survey program is to measure accurate colors of a large number of W UMa 
systems brighter than 14$^{th}$ magnitude visible from our observatory in Sonoita, Arizona. 
The equipment at the Sonoita Research Observatory (SRO) consists of a 0.35m Schmidt-Cassegrain 
telescope on a robotic German equatorial mount, a Santa Barbara Instrument Group STL-1001E CCD 
camera and a Custom Scientific $BVR_CI_C$ filter set with the \cite{bes90} prescriptions. With this equipment, we can achieve photometric precisions of 0.01 magnitudes at 14$^{th}$ magnitude with reasonable integration times. 

The observatory sits on a private site located about 60 miles east-southeast of Kitt Peak National
Observatory in very dark skies. Previously unknown main-belt asteroids in the 19$^{th}$-20$^{th}$ 
magnitude range have been discovered in five-minute unfiltered images taken at SRO, and stacked, 
unfiltered images have revealed 23$^{rd}$ magnitude stars. Seeing is typically 2\arcsec-3\arcsec\ full-width 
at half maximum.

\subsection{Software}

As with all robotic observatories, the main determinant of observing efficiency at SRO has been
the software used to control the telescope and camera. All of the observing software we use is commercially available. The two key software programs are ACP Observatory Control Software (hereafter, ACP) and ACP Scheduler from DC-3 Dreams, Incorporated of Mesa, Arizona.\footnote{Web site at \url{http://www.dc3.com}} Overall operations control is handled by ACP. At the beginning of each night, it opens the dome, and takes twilight flats. During the night, it monitors the weather conditions (closing the dome if the weather degrades to a dangerous state), moves the telescope and dome, instructs the camera control software to take images, and monitors the telescope focus, refocusing as needed. At the end of the night, it takes more twilight flats and then parks the telescope and closes the dome. The entire operation is run unattended, allowing for a tremendous increase in observing efficiency over traditional human-driven telescope operations.

The primary enabler of a survey project like this one is the scheduling software. The ACP Scheduler is a real-time scheduler that can adjust to changing observing conditions and observing priorities, as opposed to the traditional queue schedulers that decide the observing plan at the beginning of the night. Executing a survey like this one involves merely entering the coordinates of the desired fields (i.e. standard star fields and the object fields), filters and exposure times into the ACP Scheduler database once. At the start of each night, the scheduler chooses the best field to observe. When that field is completed, a new field is chosen based on the current conditions. If the weather degrades from photometric conditions, but is still reasonable for differential photometry projects, the scheduler will begin doing those. Should the weather later improve, it will return to projects like the survey that require photometric skies. The scheduler thus makes the most efficient use of the observing time, all without human intervention.

\section{Data Gathering and Processing}

While the observatory operation for the survey is untraditional, the data gathering is not. The goal is to measure the program stars in four Johnson-Cousins filters ($BVR_CI_C$) and transform the observations to the standard system via the observation of Landolt (1983, 1992) standard fields. Our approach is to separate the determination of the extinction coefficients from the transformation coefficients. We observe a reasonably bright star with a $B-V$ value near zero through each filter periodically during the night as it moves through a reasonable range of airmass, typically from 1.0 to 2.5 airmasses. The star is chosen so that it can be observed low in the eastern sky at the beginning of the night, up to the zenith, and then in the western sky late in the night. This approach typically yields about a dozen measurements through each filter per night, and covers both sides of the meridian, providing a diagnostic of the photometric quality of the night. A linear fit of instrumental magnitude versus airmass then provides the (first-order) extinction coefficient.

Observations of standard and target fields are made in the sequence $I_C$,$R_C$,$V$,$B$, and then $B$,$V$,$R_C$,$I_C$ to minimize extinction issues. The two magnitude measurements for each star in each filter are averaged to provide coeval values for forming colors. Determination of the transformation coefficients is done in the usual way by observing several Landolt standard fields near the meridian during the night and then performing linear fits to the various instrumental versus standard magnitude plots.

As images are collected during the night, they are losslessly compressed and transferred to a storage server at the lead author's institution. Processing of the images is performed with scripts written in PyRAF, a Python interface to IRAF available from the Space Science Institute web site.\footnote{PyRAF is available at \url{http://www.stsci.edu/resources/software_hardware/pyraf}} They are then bias and dark subtracted and flatfielded. Extinction coefficients are first determined and then the transformation coefficients determined. This part of the process is interactive, with plots that allow the user to determine the quality of the night and delete obviously bad images from the reduction. Once the extinction and transformation coefficients are known, the images of target fields can be reduced and the observations transformed onto the standard system. Aperture photometry is performed with an aperture radius of eight pixels, sky annulus radius of twelve pixels and an annulus thickness of three pixels. The pixels subtend 1\arcsec.24  on the sky.

\section{Survey Targets and Results}

The survey began in 2005 with targets selected from the \cite{prib03}, hereafter PKT, catalog of 361 overcontact systems. Soon thereafter, \cite{get06}, hereafter GGM, published a catalog of 1022 overcontact systems identified from ROTSE-I sky patrol images. The final list of observed targets came primarily from these two catalogs, with two systems (GSC 0067-00348 and GSC 2545-00970) added after we made follow-up observations of new discoveries. The last of the observations reported herein were made in 2008. Observations were made on 237 photometric nights, with this project sharing telescope time with several other survey projects.

In total, 606 systems with $9\le V_{max}\le13$ had observations of sufficient quality to be listed in Table 1. The table is given in abbreviated form in the printed version of the paper to show the file structure, with the full table given in the online version. The star identifications are either the GCVS designation, if known, or the designation given in GGM. The objects listed by their indentification from the GGM catalog can be searched in the SIMBAD database by using the "ggm2006" catalog identifier. We chose to retain the GGM designation because SIMBAD does not list Guide Star Catalog identifications of many of these systems. The positions, orbital periods and maximum/minimum $V$ magnitudes are taken from either the PKT or the GGM catalogs. In a handful of cases, the positions are from astrometry of our images where the GGM positions were ambiguous due to blending. All systems were observed on at least three nights but 251 systems had observations on four or more nights. The $B-V$, $V-R_C$, and $R_C-I_C$ colors are given as measured, and the $V-I_C$ color is formed as the sum of $V-R_C$ and $R_C-I_C$. The errors given are the standard deviations of the individual measurements, and this include both the measurement errors and any intrinsic color changes in the stars, which we assume to be small.

\begin{figure}
\epsscale{1.0}
\plotone{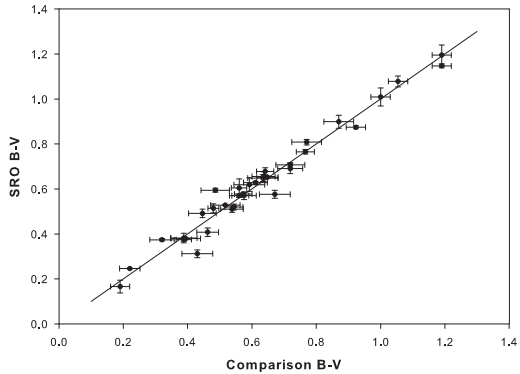}
\caption{SRO $B-V$ values compared to $B-V$ values in the literature for 35 systems.\label{fig1}}
\end{figure}

\begin{figure}
\epsscale{1.0}
\plotone{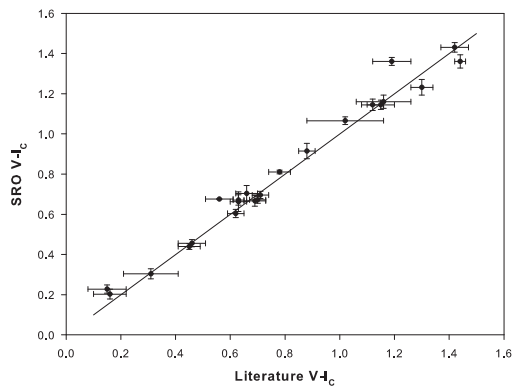}
\caption{SRO $V-I_C$ values compared to $V-I_C$ values in the literature for 23 systems.\label{fig2}}
\end{figure}

To assess the accuracy of the colors, they were compared to values found in the Tycho catalog and from other sources in the literature. The bulk of the comparison data came from the Tycho catalog and were selected for comparison to our data if the error listed in the Tycho catalog was 0.05 mag or less. To extend the comparison past $B-V=1$, values for CC Com \citep{bradstreet85}, VZ Psc \citep{bradstreet85}, and V523 Cas \citep{sfw04} were taken from the literature. Figure \ref{fig1} shows our $B-V$ values plotted against the comparison data. Figure \ref{fig2} shows our $V-I_C$ values plotted against values from Tycho and the literature. Once again, the red end of the figure was extended past the range of accurate Tycho values, with the value for XY Leo taken from \cite{hrivnak85} and values for CC Com, VZ Psc, V523 Cas, V574 Lyr, CT Tau, LR Cam, 766402, 7710246, and 13149917 taken from The Amateur Sky Survey Mark IV (TASS) \citep{droege06}. On the blue end, values for CT Tau and 2962451 were also taken from the TASS catalog. Both figures show good agreement between our photometry and that of others.

\begin{figure}
\epsscale{1.0}
\plotone{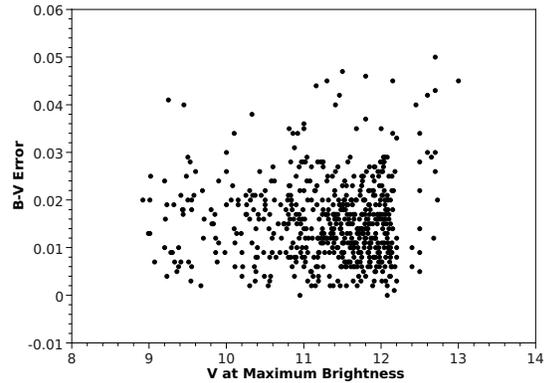}
\caption{The error in $B-V$ versus the $V$ magnitude for all systems in the survey.\label{fig3}}
\end{figure}

We examined our data for any systematic dependences of the errors on various quantities. One obvious dependence would be on the magnitude of the binary. In a constant-exposure survey, this type of dependence is expected and easily revealed. Our survey, however, had exposures optimized for each individual system, and our targets are all in the brightness range where we expect intrinsic color variability and the transformation to the standard system to be the dominant noise sources, rather than sources like photon noise or sky background noise. In that case, we do not expect to see any strong dependence on the brightness of the system and Figure \ref{fig3} shows this to be the case. 

\begin{figure}
\epsscale{1.0}
\plotone{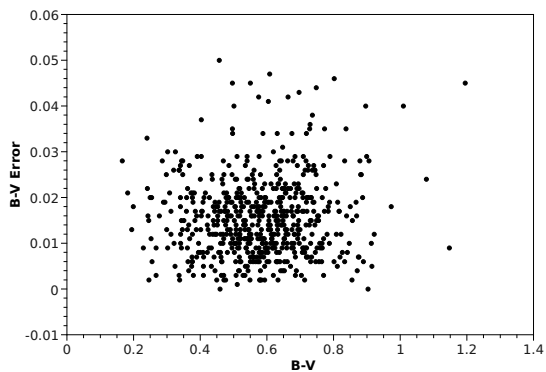}
\caption{The error in $B-V$ versus $B-V$ for all systems in the survey.\label{fig4}}
\end{figure}

Another possibility is a dependence of the color error on the color itself. Our exposure times were determined using the V magnitude of the binary at maximum brightness and the known sensitivity of the equipment for each filter. Redder systems would require a longer exposure in $B$ than a bluer system of the same brightness in order to have the same signal-to-noise ratio for photon noise (the dominant noise source in a given exposure for our equipment and the magnitude range of our survey). Since our exposures were set to achieve a signal-to-noise ratio of 200 or better (for photon noise) for each exposure, we still expect the observed errors to be dominated by intrinsic variablity and transformation uncertainties. Figure \ref{fig4} shows that this is indeed the case.

\section{Reddening}

To be of the most usefulness, the colors of the stars need to be corrected for the effect of interstellar reddening. Normally, an accurate measurement of the reddening requires a detailed spectroscopic analysis, such as that employed by \cite{terrell03} using the Na D lines as calibrated by \cite{munari97}. For the majority of the systems in our survey, such data do not exist. We therefore present a preliminary analysis using two approaches for which complete data are available: determining a reddening-free quantity from our photometry, and computing the reddening from the neutral hydrogen column density, $N_{HI}$, along the line of sight to the system. The former approach, referred to hereafter as the color difference method, follows that used for early-type stars with $U-B$ and $B-V$ colors to determine the reddening-free parameter $Q$ \citep{johnsonmorgan53}. The latter is the method employed by \cite{rucinski97} in calibrating the absolute magnitudes of W UMa systems. While these approaches have limitations, as discussed below, we believe they can provide useful information until better data can be obtained.

Following conventional notation, we write the observed color as $B-V$, the unreddened color as $(B-V)_0$, and the reddening as $E(B-V)$ so that $B-V = (B-V)_0 + E(B-V)$. The color difference method uses the difference in two colors to form a reddening-free quantity usually respresented by the letter $Q$. The key is to mutiply one of the colors by a constant factor representing the ratio of the color excesses for the two colors, \it i.e.\rm, $Q=C1-\frac{(E(C1)}{E(C2)}\times C2$. While the traditional $Q$ quantity, $(U-B)-\frac{E(U-B)}{E(B-V)} \times (B-V)$, is not useful for the stars of spectral types in this survey, the basic idea can be applied to more appropriate filters. Here we use the $B-V$ and $V-I_C$ colors, and thus define $Q$ as $(B-V)-\frac{E(B-V)}{E(V-I_C)} \times (V-I_C)$. Using the absorption ratios, $\frac{A(\lambda )}{A(V)}$, given in \cite{cardelli89} and assuming a total to selective extinction ratio of $R_V=3.1$, the ratio $\frac{E(B-V)}{E(V-I)}$ has a value of 0.647. 

\begin{figure*}
\epsscale{1.9}
\plotone{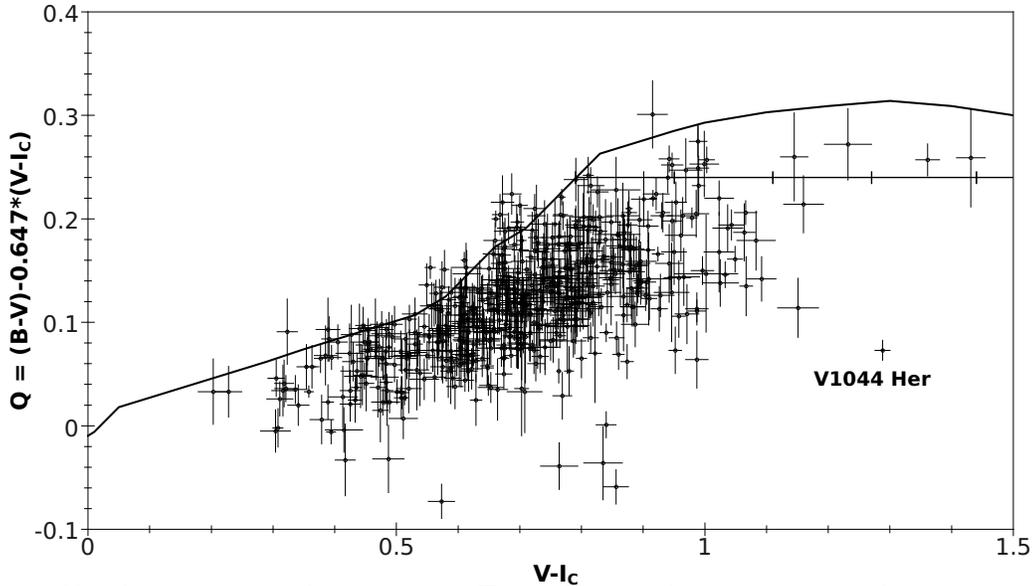}
\caption{$Q$ versus $V-I_C$ for the stars in the survey. The solid curve is the zero-age main sequence from \cite{walker85} and the horizontal line is a reddening line for a star with $(V-I_C)_0=0.79$ with the ticks representing steps of 0.1 in $E(B-V)$.\label{fig5}}
\end{figure*}

Figure \ref{fig3} is the color difference diagram for all of the systems in the survey. The solid curve shows the main sequence from \cite{walker85} for $V-I_C\le 0.95$ and \cite{cousins78} for $V-I_C > 0.95$. Because $Q$ is reddening-independent, the reddening lines in such a diagram are horizontal and can be used to estimate the reddening of a given star. The horizontal line in Figure \ref{fig5} shows the reddening of a main sequence star with $(V-I_C)_0=0.79$ and the vertical ticks indicate increasing reddening in steps of 0.1 in $E(B-V)$. 

One concern with this approach to determining the reddening is that many, perhaps all \citep{prib08}, of these objects have companions and this can have a large effect on the color difference diagram. If, as is usually the case, the companions are cool, low-mass objects, they will affect the longer wavelength filters more than the shorter ones, and make the systems look redder than they otherwise would be. However, except in most circumstances, we expect the third bodies to be of low luminosities, making the effect on the system colors fairly small, perhaps a few hundredths of a magnitude. Unfortunately, these errors are large enough to greatly diminish the usefulness of the color difference diagram for systems only slightly affected by interstellar reddening, as we expect most of these systems to be due to their brightnesses, and thus nearby distances.

While not as useful as we had hoped for determining reddening, the color difference diagram does distinguish a few systems as being unusual. Figure \ref{fig3} shows a few systems at large horizontal distances from the main sequence curve, such as V1044 Her whose $E(V-I_C)$ value from the color difference diagram is about 0.9. The question is whether that is caused by large interstellar reddening, or something else, such as a luminous companion. Looking at the expected reddening from the neutral hydrogen column density provides a clue. 

\cite{predehl95} used ROSAT observations to determine the relationship between X-ray absorption and optical extinction. Again assuming  $R_V=3.1$, their results show that $E(B-V)=1.8\times 10^{-22}\times N_{HI}$. To compute the $N_{HI}$ value for individual systems, we used the ``nh'' tool in the ``heasarc'' sub-package of the FTOOLS software available from the NASA Goddard Spaceflight Center.\footnote{Available at \url{http://heasarc.gsfc.nasa.gov/lheasoft/ftools/heasarc.html}} Although this approach to estimating the reddening doesn't suffer the same problems as the color difference method, it does have potential problems. The main issue is that the $N_{HI}$ column density is estimated by interpolating between the approximately 0.5 degree sampling points of the $N_{HI}$ surveys as described in the software documentation. Caution must be exercised in using this method in areas where there are inhomogeneities in the interstellar medium on the scale of the interpolation. With that caveat in mind, we can compute the reddening and compare it to the value from the color difference method. 

In the case of V1044 Her, we find that the value of $E(B-V)$ from estimated from the $N_{HI}$ column density is 0.05, while $E(B-V)$ from the color difference method is 0.58. In order for the color difference value to be correct, the system would have to have a $(B-V)_0$ value of about 0.34, making it an early F-type star, but its orbital period of about 0.24 days precludes a spectral type that early. In this case, the reddening estimate based on the $N_{HI}$ column density seems much more reasonable, but that leaves unanswered the question of why the $B-V$ value, implying $T_{eff}\approx5000K$, and the $V-I_C$ value, implying $T_{eff}\approx 4300K$, are so disparate. The $V-R_C$ and $R-I_C$ values, 0.69 and 0.60 respectively, are consistent with the $V-I_C$ value, implying that the $B-V$ value is too blue. The system was measured on three nights within a week of one another, and the three $B-V$ values were consistent with one another, with values of 0.91, 0.92 and 0.92. The images show no obvious problems and the unfiltered light curve of \cite{blatt01} does not appear to have large amounts of third light, so the reason for the disparity remains unclear. Perhaps the system is more active than usual, leading to elevated brightness in the $B$ filter. Being one of only a few known systems with an orbital period less than 0.25 days, further observational study of V1044 Her is warranted.

The $N_{HI}$ method occasionally produces reddening values that are very large compared to the color difference method. One example is V1191 Cyg whose $E(B-V)$ from the color difference method is 0.05 while the $N_{HI}$ method yields a value of 1.13. This difference is not very surprising since Cygnus lies along the galactic plane and large inhomogeneities in the interstellar medium are to be expected. Because of the weaknesses of each method, the safest approach is to trust only reddening values where the two methods are in good agreement. There are 321 systems that have equal values within the error bars, with the error in the $N_{HI}$ method assumed to be 0.02 magnitudes. However, of those, many have fairly large errors in the values from the color difference method, with many having errors significantly larger than the value itself. To generate a more useful list, we further culled the list by eliminating systems where the difference between the two methods was 0.02 mag or less. This culling leaves 101 systems where the reddening values between the two methods are consistent. Table 2 shows how the values compare for these 101 systems.

\begin{deluxetable}{rrrlcccccrr}
\tablewidth{7.05in}
\setlength{\tabcolsep}{0.05in} 
\tabletypesize{\scriptsize}
\tablecaption{The $BVR_CI_C$ colors of 606 W Ursae Majoris binaries.}
\tablenum{1}
\tablehead{\colhead{System} & \colhead{R.A.} & \colhead{Decl.} & \colhead{Period} & \colhead{Num.} & \colhead{$B-V$} & \colhead{$V-R_C$} & \colhead{$R_C-I_C$} & \colhead{$V-I_C$} & \colhead{$V_{max}$} & \colhead{$V_{min}$} \\ 
\colhead{} & \colhead{($^{\circ}$ J2000)} & \colhead{($^{\circ}$ J2000)} & \colhead{(days)} & \colhead{Obs.} & \colhead{} & \colhead{} & \colhead{} & \colhead{} & \colhead{} & \colhead{}} 
\startdata
299755 & 1.4917 & 78.9081 & 0.43821290 & 3 & $0.562\pm0.016$& $0.332\pm0.003$ & $0.348\pm0.004$ & $0.680\pm0.005$ & 11.83 & 12.50 \\
1557554 & 1.6542 & 55.4561 & 0.27249130 & 3 & $0.874\pm0.007$ & $0.510\pm0.007$ & $0.432\pm0.016$ & $0.942\pm0.017$ & 9.33 & 9.64 \\
V1007 Cas & 2.0125 & 51.1342 & 0.33201060 & 3 & $0.724\pm0.027$ & $0.459\pm0.004$ & $0.437\pm0.010$ & $0.896\pm0.011$ & 12.03 & 12.43 \\
11936075 & 3.9833 & 6.7457 & 0.40117920 & 3 & $0.519\pm0.012$ & $0.308\pm0.008$ & $0.332\pm0.009$ & $0.640\pm0.012$ & 11.30 & 11.71 \\
9107370 & 4.0583 & 22.1661 & 0.42090500 & 3 & $0.512\pm0.009$ & $0.324\pm0.015$ & $0.322\pm0.014$ & $0.646\pm0.021$ & 11.49 & 11.82 \\
CN And & 5.1260 & 40.2261 & 0.46279111 & 4 & $0.516\pm0.012$ & $0.277\pm0.011$ & $0.278\pm0.014$ & $0.555\pm0.018$ & 9.70 & 10.25 \\
6335770 & 6.3292 & 32.4657 & 0.54349740 & 3 & $0.377\pm0.021$ & $0.231\pm0.006$ & $0.236\pm0.004$ & $0.467\pm0.007$ & 10.25 & 10.56 \\
9093368 & 6.7958 & 10.0151 & 0.52122360 & 4 & $0.534\pm0.013$ & $0.313\pm0.011$ & $0.299\pm0.014$ & $0.612\pm0.018$ & 10.19 & 10.61 \\
14680632 & 7.0875 & -14.8881 & 0.40264980 & 3 & $0.577\pm0.017$ & $0.351\pm0.007$ & $0.351\pm0.017$ & $0.702\pm0.018$ & 11.56 & 11.91 \\
254036 & 7.1167 & 78.9619 & 0.31741200 & 3 & $0.586\pm0.008$ & $0.356\pm0.015$ & $0.350\pm0.005$ & $0.706\pm0.016$ & 10.56 & 10.93 \\
CL Cet & 7.2667 & -17.2169 & 0.62161800 & 3 & $0.389\pm0.007$ & $0.221\pm0.013$ & $0.233\pm0.007$ & $0.454\pm0.015$ & 9.88 & 10.00 \\
3667194 & 7.6417 & 39.4564 & 0.41711880 & 3 & $0.571\pm0.018$ & $0.337\pm0.003$ & $0.354\pm0.002$ & $0.691\pm0.004$ & 12.08 & 12.47 \\
9141373 & 7.6917 & 14.5461 & 0.34255000 & 4 & $0.589\pm0.011$ & $0.384\pm0.017$ & $0.345\pm0.017$ & $0.729\pm0.024$ & 11.22 & 11.55 \\
9124077 & 8.5542 & 20.8739 & 0.34014730 & 3 & $0.596\pm0.008$ & $0.392\pm0.017$ & $0.368\pm0.001$ & $0.760\pm0.017$ & 11.91 & 12.21 \\
\enddata
\tablecomments{Table 1 is published in its entirety in the electronic edition of the Astronomical Journal. A portion is shown here for guidance regarding its form and content.}
\end{deluxetable}

\begin{figure*}
\epsscale{1.95}
\plotone{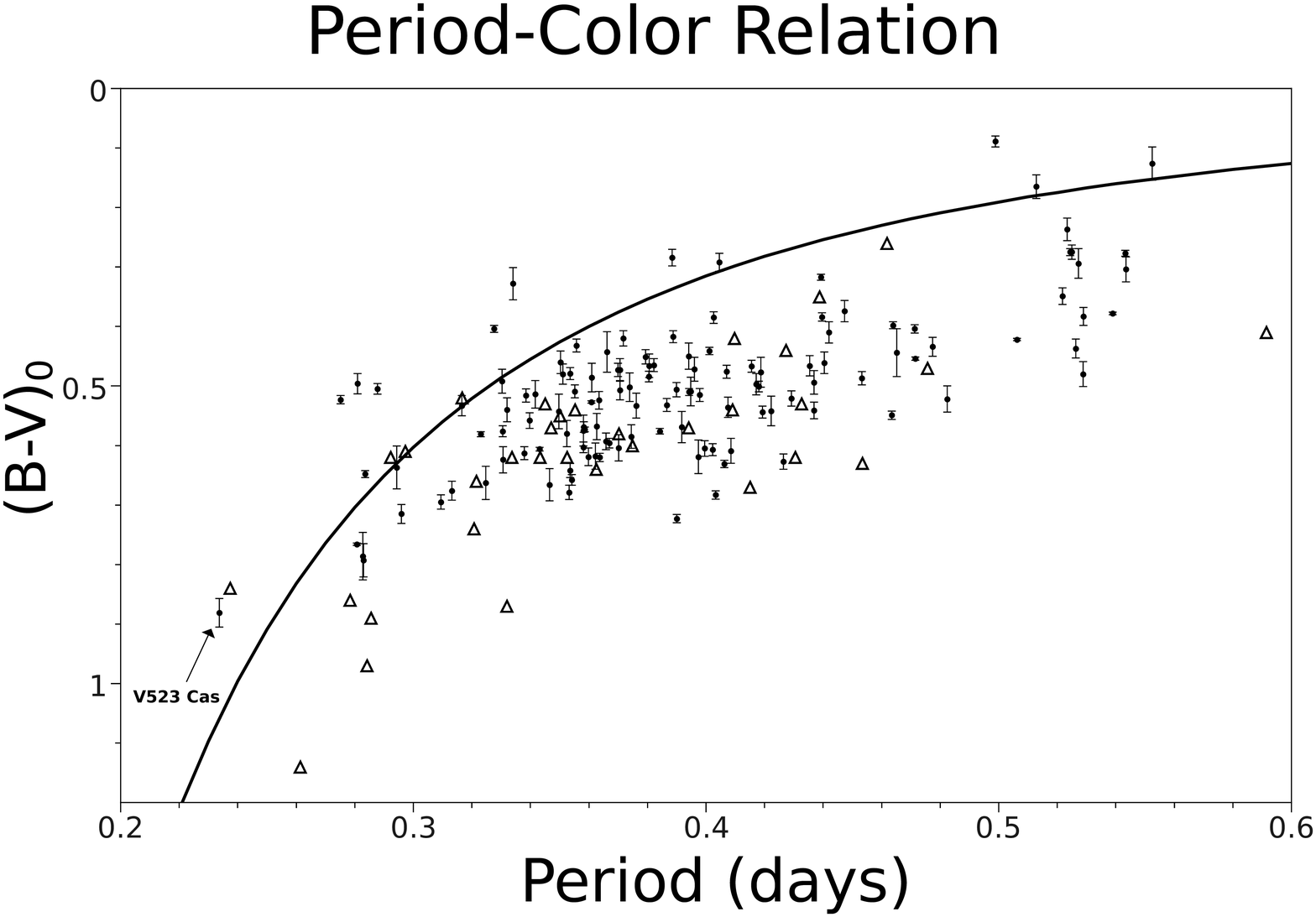}
\caption{Period-color relation for the systems in Table 2 (with error bars) and the systems from \cite{rucinski97} with periods less than 0.6 days (triangles). The solid curve is the short period blue envelope (SPBE) from \cite{rucinski98}\label{fig6}. V523 Cas is one of several systems bluer than the SPBE and is discussed in the text.}
\end{figure*}

Figure \ref{fig6} shows the period-color relation for the systems in Table 2 along with the systems from \cite{rucinski97} with periods less than 0.6 days. The solid line is the short period blue envelope (SPBE) from \cite{rucinski98} which can be likened to the zero-age main sequence for overcontact binaries in that systems will evolve towards redder colors and longer periods. Overcontact systems are thus not expected to be bluer than the SPBE. The systems in Figure \ref{fig5} that are bluer than the SPBE by more than one standard deviation in the $B-V$ value (3091792, 6133549, 3947987, 2569021, 9649548, 3962973, 8299111, 6225358, and V523 Cas) could be systems that are not overcontact systems, systems with a blue companion or they may be systems where our estimate of the reddening is too large. In the case of V523 Cas (whose overcontact nature appears solidly established), \cite{sfw04} find $B-V=1.03$ to 1.09 for the binary and $(B-V)_0=0.97\pm0.04$ for the primary component, whereas we find $B-V=1.08\pm0.02$ and $(B-V)_0=0.89\pm0.03$ for the binary. Their analysis of the O-C digram results in a third body mass of about $0.4 M_{\sun}$, which would not explain the ``blue excess'' of the system, and since there is good agreement between their reddened $B-V$ and ours, the difference may lie in the determination of the reddening, or perhaps intrinsic activity in the system that enhanced the brightness in the $B$ filter during our observations. We note that both the color difference and the $N_{HI}$ methods show good agreement for the reddening in V523 Cas (see Table 2). Further observational studies of these systems will be needed to determine their nature.

\section{Conclusions}

Standardized photometry in Johnson-Cousins $BVR_CI_C$ filters was obtained for 606 known or suspected W UMa eclipsing binaries. 
\cite{get06}, from which a large fraction of the targets in this survey were taken, estimate that perhaps 5\% of the systems in their 
catalog may be objects other than W UMa systems, so we have a reasonably pure sample of such binaries. Comparison of our results with previously published photometry on a few dozen systems shows good agreement. Estimates of the interstellar reddening were made with the color difference method and by using the neutral hydrogen column density along the line of sight to the stars. While neither approach is ideal, they each have strengths and are based on independent observational data. A subset of 101 systems in the survey shows good agreement between the two methods. It is hoped that these data will be useful in further research to understand the nature of these very interesting binaries.

\begin{deluxetable}{rccrccrcc}
\tablewidth{6.75in}
\setlength{\tabcolsep}{0.02in} 
\tabletypesize{\scriptsize}
\tablecaption{Reddening estimates for the 101 systems showing a maximum difference of 0.02 mag in $E(B-V)$ between the color difference method and the $N_{HI}$ method.}
\tablenum{2}
\tablehead{\colhead{System} & \colhead{$E(B-V)_{CD}$} & \colhead{$E(B-V)_{N_{HI}}$} & \colhead{System} & \colhead{$E(B-V)_{CD}$} & \colhead{$E(B-V)_{N_{HI}}$} & \colhead{System} & \colhead{$E(B-V)_{CD}$} & \colhead{$E(B-V)_{N_{HI}}$}}
\startdata
V523 Cas&0.197$\pm$0.177&0.191&3091792&0.193$\pm$0.042&0.188&GSC 0067-00348&0.110$\pm$0.134&0.127\\
6225358&0.098$\pm$0.022&0.087&7209961&0.124$\pm$0.042&0.144&14541307&0.085$\pm$0.031&0.071\\
CW Cmi&0.076$\pm$0.042&0.077&SX Crv&0.037$\pm$0.055&0.054&HO Psc&0.054$\pm$0.075&0.054\\
3962973&0.118$\pm$0.064&0.106&5447547&0.090$\pm$0.061&0.078&11670840&0.098$\pm$0.033&0.087\\
DF Hya&0.030$\pm$0.055&0.047&V1007 Cas&0.184$\pm$0.045&0.171&3947987&0.207$\pm$0.135&0.199\\
76033&0.079$\pm$0.046&0.081&13007821&0.046$\pm$0.042&0.039&4666562&0.097$\pm$0.033&0.098\\
AZ Vir&0.033$\pm$0.078&0.025&LZ Dra&0.108$\pm$0.082&0.099&15697139&0.041$\pm$0.053&0.057\\
AC Boo&0.039$\pm$0.058&0.021&10724466&0.056$\pm$0.032&0.065&13120541&0.052$\pm$0.046&0.037\\
14819714&0.064$\pm$0.030&0.080&14540910&0.061$\pm$0.033&0.071&CK Boo&0.011$\pm$0.033&0.025\\
4701979&0.109$\pm$0.051&0.106&QW Gem&0.161$\pm$0.064&0.145&V829 Her&0.042$\pm$0.025&0.022\\
VZ Lib&0.090$\pm$0.049&0.107&16148635&0.074$\pm$0.017&0.085&5269546&0.008$\pm$0.042&0.017\\
V1097 Her&0.052$\pm$0.023&0.059&AW CrB&0.019$\pm$0.080&0.018&12209034&0.131$\pm$0.056&0.144\\
4750697&0.095$\pm$0.049&0.079&7329879&0.052$\pm$0.040&0.060&11899139&0.065$\pm$0.024&0.058\\
DZ Psc&0.054$\pm$0.130&0.041&GSC 2545-00970&0.000$\pm$0.033&0.016&12049671&0.048$\pm$0.046&0.044\\
5261236&0.024$\pm$0.057&0.025&XY Boo&0.011$\pm$0.043&0.026&14378336&0.070$\pm$0.079&0.072\\
17385764&0.038$\pm$0.066&0.028&HV Aqr&0.059$\pm$0.059&0.062&7493998&0.037$\pm$0.060&0.025\\
1033806&0.055$\pm$0.059&0.041&LO And&0.140$\pm$0.031&0.155&11370982&0.110$\pm$0.079&0.130\\
7269084&0.086$\pm$0.056&0.093&BI CVn&0.017$\pm$0.021&0.018&10142767&0.041$\pm$0.030&0.054\\
5214166&0.041$\pm$0.036&0.026&V1036 Her&0.069$\pm$0.061&0.059&7336416&0.068$\pm$0.092&0.065\\
17357108&0.048$\pm$0.021&0.047&609692&0.131$\pm$0.095&0.137&2643685&0.013$\pm$0.063&0.027\\
AR CrB&0.043$\pm$0.069&0.050&908512&0.070$\pm$0.046&0.084&9221283&0.074$\pm$0.033&0.067\\
BL Dra&0.083$\pm$0.079&0.073&14591442&0.051$\pm$0.032&0.048&6099330&0.178$\pm$0.089&0.172\\
10476945&0.008$\pm$0.022&0.026&2888729&0.041$\pm$0.046&0.035&4805949&0.048$\pm$0.060&0.056\\
2657877&0.024$\pm$0.035&0.021&3667194&0.074$\pm$0.045&0.075&15779695&0.079$\pm$0.081&0.091\\
10608583&0.053$\pm$0.038&0.055&4873888&0.038$\pm$0.059&0.020&4760150&0.072$\pm$0.059&0.059\\
2795010&0.023$\pm$0.048&0.023&13259318&0.042$\pm$0.059&0.028&13363862&0.032$\pm$0.033&0.050\\
XY LMi&0.021$\pm$0.060&0.021&ET Psc&0.057$\pm$0.040&0.061&6375021&0.078$\pm$0.046&0.079\\
16134469&0.074$\pm$0.094&0.093&11905222&0.041$\pm$0.023&0.046&9311873&0.241$\pm$0.041&0.228\\
15595338&0.057$\pm$0.142&0.057&V728 Her&0.006$\pm$0.046&0.024&14560232&0.059$\pm$0.069&0.071\\
V2759 Ori&0.254$\pm$0.047&0.261&10141474&0.054$\pm$0.042&0.056&14304024&0.084$\pm$0.104&0.097\\
1206915&0.102$\pm$0.076&0.101&1033721&0.057$\pm$0.093&0.039&3791537&0.101$\pm$0.052&0.116\\
RU UMi&0.037$\pm$0.062&0.019&V843 Mon&0.241$\pm$0.083&0.224&BC Eri&0.053$\pm$0.130&0.072\\
TZ Lyr&0.043$\pm$0.066&0.052&ES UMa&0.056$\pm$0.079&0.073&14678045&0.047$\pm$0.052&0.044\\
6335770&0.073$\pm$0.093&0.070&2962451&0.040$\pm$0.127&0.043\\
\enddata
\end{deluxetable}

\acknowledgments
We thank the referee, Slavek Rucinski, for a helpful review that greatly improved the quality of the paper. This research has made use of the SIMBAD database, operated at CDS, Strasbourg, France.

\end{document}